\documentclass{article}
\usepackage[T1]{fontenc} 
\usepackage[utf8]{inputenc} 
\usepackage{ismir,amsmath,cite,url}
\usepackage{graphicx}
\usepackage{color}
\usepackage{subcaption}

\title{Real-Time Percussive Technique Recognition
and Embedding Learning for the Acoustic Guitar}

\threeauthors
  {Andrea Martelloni} {Queen Mary University \\ of London \\ \small{{\tt a.martelloni@qmul.ac.uk}}}
  {Andrew P McPherson} {Imperial College \\ \\ \small{{\tt andrew.mcpherson@imperial.ac.uk}}}
  {Mathieu Barthet} {Queen Mary University \\ of London \\ \small{{\tt m.barthet@qmul.ac.uk}}}

\def\authorname{A. Martelloni, A. P. McPherson, M. Barthet}

\begin{document}

\setlength{\abovedisplayskip}{0pt}
\setlength{\belowdisplayskip}{0pt}

\maketitle
\begin{abstract}


Real-time music information retrieval (RT-MIR) has much potential
to augment the capabilities of traditional acoustic instruments.
We develop RT-MIR techniques aimed at augmenting percussive fingerstyle,
which blends acoustic guitar playing with guitar body percussion.
We formulate several design objectives for RT-MIR systems for
augmented instrument performance: (i) causal constraint,
(ii) perceptually negligible action-to-sound latency,
(iii) control intimacy support, (iv) synthesis control support.
We present and evaluate real-time guitar body percussion recognition
and embedding learning techniques based on convolutional neural networks
(CNNs) and CNNs jointly trained with variational autoencoders (VAEs).
We introduce a taxonomy
of guitar body percussion based on hand part and location.
We follow a cross-dataset evaluation approach by collecting
three datasets labelled according to the taxonomy. The embedding
quality of the models is assessed using KL-Divergence across
distributions corresponding to different taxonomic classes.
Results indicate that the networks are strong classifiers especially
in a simplified 2-class recognition task, and the VAEs yield
improved class separation compared to CNNs as evidenced
by increased KL-Divergence across distributions.
We argue that the VAE embedding quality could
support control intimacy and rich interaction when 
the latent space's parameters are used to control an
external synthesis engine.
Further design challenges
around generalisation to different datasets
have been identified.

\end{abstract}
\section{Introduction}\label{sec:introduction}

There is increasing interest in deep neural networks
for processing audio in real time with sufficiently
low latency to be used in musical performance.
There is also a drive to provide small self-contained platforms
that could perform inference \emph{at the edge},
that is, on a device that
can be fitted in a musical interface or a musical
instrument
\cite{turchet2018internet,Pelinski2022Embedded,stefani2022comparison}.
Many of the tasks in Music Information Retrieval, such as
onset detection \cite{bock2012online},
playing technique classification \cite{wang2020spectral},
timbre transfer \cite{jain2020att},
re-synthesis of musical information \cite{caspeDDX7DifferentiableFM2022}
and generative composition \cite{liang2017automatic},
find an application in the design of
Digital Musical Instruments (DMI) and augmented instruments,
as long as the solutions conform to real-time requirements.
For Real-Time MIR (RT-MIR), two physical constraints
that limit the application of Deep Neural Network (DNN)
models are \emph{causality}, implying the inability to look into
the future, and \emph{low action-to-sound latency}
\cite{stefaniChallengesEmbeddedRealtime2022}.
Acceptable action-to-sound latency in music performance
was found to be 10 ms \cite{wesselProblemsProspectsIntimate2002}
for percussion instruments, and
the latency's \emph{jitter} (the variation)
was also found to be a factor in the quality of the
interaction \cite{McPherson2016}.
Although there are ways
to work around higher latencies, for example by
synthesising generic attacks before a specific sound
is generated \cite{stowellDelayedDecisionmakingRealtime2010},
the ideal approach would be to develop a system fulfilling
the latency constraints in the first place.

In this work, we investigate RT-MIR for the processing
and mapping of guitar body hit sounds to augment the
timbral palette of the instrument in percussive fingerstyle.
Percussive
fingerstyle is an extended guitar technique that uses
layered arrangements, alternate tunings and hits on the
guitar's body to create the impression of a
``one-man band'' \cite{NIME20_85}.
Our method relies on deep learning to develop recognition
and embedding learning
of guitar body percussion. Our model addresses the task
of generating representations of body hits according to
performers' percussive gestures, separating them by hand part and
location.
One possible application is
to map such a description as parameters for a synthesis engine,
such as real-time physics-based synthesis.
We
adapted an Automatic Drum Transcription (ADT)
model based on a Convolutional Neural Network (CNN)
to fit the practical
constraints of an augmented instrument for percussion.
Our longer-term
aim is to design a network that not only works as a classifier,
but also describes guitar body hits
with a set of features unique for each sample, to support
control intimacy \cite{mooreDysfunctionsMIDI1988}
and try to achieve the same level of nuance
afforded by acoustic instruments. To this end, we propose a
variation of our model that jointly trains a classifier and
a Variational Autoencoder (VAE)
\cite{kingmaAutoEncodingVariationalBayes2014}.

\section{Background}\label{sec:background}

\textbf{Percussion DMIs.}
In opposition to the direct
control offered by acoustic percussion,
digital percussion instruments have historically afforded indirect
control of discrete events \cite{wanderley2001gestural},
with hit dynamics often being
the only expressive parameter over individual hits.
Jathal \cite{jathalRealTimeTimbreClassification2017} provides
a detailed description of commercial digital
percussion instruments, emphasising the fact that they
force the player to adapt their technique to the tool, usually a
set of buttons or a zone-based sample trigger. The author also
advocates for the design of interfaces that interpret the
technique that performers of a particular acoustic instrument
have already mastered: traditional techniques will be the
first sensorimotor reference that expert players will use for
the exploration of DMIs \cite{jackRichGestureReduced2017}, and they have been
used in the past as the basis for controllers to navigate synthesiser
spaces \cite{tremblaySurfingWavesLive2010}.

\textbf{Hit classification.}
One approach is to take data from audio transducers
or other sensors
for on-the-fly event detection and classification, and
the use of that data to trigger the generation of a
sound associated to that category. Examples are Turchet \emph{et al.}'s
Smart Caj\'on \cite{turchetRealTimeHitClassification2018}, Jathal's
HandSolo \cite{jathalRealTimeTimbreClassification2017} and
Zamborlin \emph{et al.}'s Mogees \cite{zamborlinStudiesCustomisationdrivenDigital}.
This approach is well supported by music tools and software for
machine learning in music such as \texttt{bonk\~} for Max/PD
\cite{pucketteRealtimeAudioAnalysis1998}, timbreID's
\texttt{barkSpec\~} and the Wekinator \cite{fiebrink2010wekinator}.
This has also been applied to the acoustic guitar through the work
of L\"ahdeoja \cite{lahdeojaAugmentingChordophonesHybrid2009}
and Stefani \emph{et al.} \cite{stefaniDemoTimbreIDVSTPlugin2020, stefani2022comparison},
the latter applying fully-connected DNN layers
for multi-class classification of guitar techniques, including percussive
ones. No direct attempts have been made to use machine learning
to achieve a description beyond classification, especially
one that would support Moore's \emph{control intimacy}
\cite{mooreDysfunctionsMIDI1988}.

\textbf{Automatic Drum Transcription in MIR.}
A task related to guitar percussion classification in MIR is Automatic
Drum Transcription, the audio-based detection and inference of score
notations for percussive parts. Current literature does not
only address the Western drum kit, but also other percussion
instruments such as the tabla \cite{narangACOUSTICFEATURESDETERMINING2017}.
A recent review \cite{wuReviewAutomaticDrum2018} reports that,
in the current state of the art in ADT, solutions either use non-negative matrix factorisation (NMF)
or
look into the relationships between hits
and tackle the problem with a language model or a recurrent neural network.
The most relevant system for our application is based on a CNN that jointly performs
event detection and
classification for ADT using a sliding buffer of 150 ms
\cite{jacquesAutomaticDrumTranscription2018} and was trained on the
MIREX17 drums dataset \cite{vogl2018mirex}. Mattur
Ananthanarayana (MA) \emph{et al.}
fine-tuned the model on a dataset of tabla strokes,
noting a resemblance between Kick Drum, Snare Drum and Hi-Hat
sounds and the tabla strokes themselves
\cite{mafour}.


\textbf{Real-time DNNs for music performance.}
Our purpose is not the direct re-synthesis of guitar body hits,
but rather the control of the parameters of a synthesis engine.
However, we are inspired by the introduction of neural networks
as tools for music performance, through solutions such as
Neural Audio Synthesis and Neural or Differentiable
Digital Signal Processing (DDSP).
Bottlenecks and latent spaces have been used with
VAEs \cite{eslingGenerativeTimbreSpaces2018, bittonModulatedVariationalAutoEncoders2018}
and autoencoder-like structures
\cite{engelDDSPDifferentiableDigital2020}
for re-synthesis of sounds and timbre transfer,
with successful real-time implementations
such as RAVE \cite{caillonRAVEVariationalAutoencoder2021}
and DDX7 \cite{caspeDDX7DifferentiableFM2022}.
NN-based solutions have also been applied to model
linear \cite{colonelDirectDesignBiquad2021} and non-linear
audio systems, such as guitar amplifiers \cite{eichasBlockorientedGrayBox2017}
and stomp-box overdrives \cite{colonelReverseEngineeringMemoryless2022, damskaggRealTimeModelingAudio2019}.
Solutions exist to load an arbitrary neural DSP network into a
plugin to be run in a DAW, such as the Neutone VST
host by
Qosmo\footnote{\url{https://neutone.space/plugin/}}
and IRCAM's
\texttt{nn\~}\footnote{\url{https://github.com/acids-ircam/nn_tilde}}
Max/PD external.


\begin{table}
  \begin{center}
  \begin{tabular}{|l|l|}
   \hline
   \textbf{Tool} & \textbf{Latency} \\
   \hline
   \texttt{bonk\~} (Puckette \cite{pucketteRealtimeAudioAnalysis1998})  & 6 ms \\
   \hline
   Stefani \emph{et al.} \cite{stefani2022comparison} & 20 ms \\
   \hline
   RAVE (Caillon \emph{et al.} \cite{caillonRAVEVariationalAutoencoder2021}) & DAW-defined \\
   \hline
   Mogees (Zamborlin \cite{zamborlinStudiesCustomisationdrivenDigital}) & 23 ms \\
   \hline
   HandSolo (Jathal \cite{jathalRealTimeTimbreClassification2017}) & 17 ms \\
   \hline
   Tabla stroke classifier (MA \emph{et al.} \cite{mafour}) & 150 ms \\
   \hline
  \end{tabular}
 \end{center}
  \caption{Reported buffer values for detection and inference for some of the
  works and studies cited in this section.}
  \label{tab:latency}
 \end{table}

 \textbf{Latency.} The impact of latency and jitter in music performance systems
was investigated, for example, by McPherson \emph{et al.} \cite{McPherson2016}.
Table \ref{tab:latency}
reports the measurements published by the authors cited so far
on the latency of their tools, specifically the duration
of analysis and inference rather than audio
input-to-output latency, which is system-dependent more
than algorithm-dependent.
Most tools that are meant for real-time use achieve latencies in
the region of 20 ms, which exceeds Wessel and Wright's
10 ms ceiling for musical instruments
\cite{wesselProblemsProspectsIntimate2002}.

\textbf{Challenges in rich representation.}
Gesture classification toolkits like \texttt{bonk\~}
have been deployed
in many music-making interfaces,
including guitars \cite{lahdeojaAugmentingChordophonesHybrid2009},
but they were shown to make percussive guitar
performers uneasy owing to the chance of misclassification
for ambiguous or unexpected inputs \cite{NIME21_32}.
Standard classifiers also do not represent subtle variability
within gestural categories,
leading to
a small gestural
\emph{bottleneck} \cite{jackRichGestureReduced2017}.
Related studies in Human-Computer Interaction (HCI) also promote the
design of DMIs sensitive to the \emph{micro-scale}
of musical actions, the scale of differences across
gestures of the same category
\cite{armitageStudyingSubtleDetailed2023}; authors have suggested
that rich and controllable behaviour could be as important
as high classification accuracy for
creative applications
\cite{vigliensoniSmallDataMindsetGenerative2022}.
Dimensionality reduction of input representations
through VAEs, as performed for example by RAVE,
could help investigate these rich dimensions.

\section{Methodology}\label{sec:methodology}


\subsection{From taxonomy to datasets}

This work builds upon our two prior studies on the
investigation of the technique of percussive fingerstyle
\cite{NIME20_85} and the design of a prototype augmented
guitar to optimally capture those techniques \cite{NIME21_32}.
Those observations firstly led to the creation of a
\emph{taxonomy of guitar body percussion},
inspired by the work by
Goddard on the taxonomy of bass playing techniques
\cite{goddardVirtuosityComputationallyCreative2021}:

\[
    {\begin{cases}
        \text{hit}\\
        \text{scrape}
    \end{cases}}
    \text{the guitar with} \;
    {\begin{cases}
        \text{heel}\\
        \text{thumb}\\
        \text{fingers}\\
        \text{nails}\\
    \end{cases}}
    \text{at the} \;
    {\begin{cases}
        \text{soundhole}\\
        \text{upper bout}\\
        \text{lower bout}\\
        \text{upper side}\\
        \text{lower side}\\
    \end{cases}}
\] \label{eq:taxonomy}

This was used to create the labelled dataset \emph{GPercRep}
by producing 50 examples (one hit per second)
of each combination of taxonomy attributes, excluding
those that are ergonomically impossible, e.g. reaching the lower sides
of the body with the heel of the hand. This leads to an imbalanced
but ecologically valid
dataset \cite{ramyachitraImbalancedDatasetClassification2014}.
Each combination was repeated at four
dynamics levels (\emph{p}, \emph{mp}, \emph{mf}, \emph{f}).
All recordings were made by the first author on the guitar built for
\cite{NIME21_32}, which has a six-channel output made out of
one magnetic pickup and five piezo sensors on each of the locations
(soundhole, etc..., see taxonomy above). The guitar had 12-53 gauge strings
in standard tuning, muted with the left hand, and the hits were played
with the bare right hand.
After excluding scrapes from the analysis, as they require a
time-based gesture follower,
the dataset has 3,157 examples extracted from 52 minutes and 37 seconds of
audio at 44.1 kHz. The dataset is currently not public.

\begin{table}\small
  \centering
  \setlength\tabcolsep{1.8pt}
    \begin{tabular}{|l|p{7em}|p{5.5em}|p{7em}|}
    \hline
    \multicolumn{1}{|r|}{} & \textbf{Input Features} & \textbf{CNN Type} & \textbf{Bottleneck} \\
    \hline
    \textit{TablaCNN} & 80-band Mel & 2-layer 2D. Kernel size: 1x7, 1x3 & 128 dimensions, reduced through PCA \\
    \hline
    \textit{PercCNN} & 512-bin FFT decimated to 64 bins & 3-layer 1D. Kernel size: 6, 5, 5 & 2 dimensions \\
    \hline
    \textit{PercVAE} & 512-bin FFT decimated to 64 bins & 3-layer 1D. Kernel size: 6, 5, 5 & 2 dimensions (\(\mu\) + \(\sigma\)) \\
    \hline
    \end{tabular}%
    \caption{Differences across network architectures used.}
  \label{tab:netarch}%
\end{table}%

\begin{table}\small
  \centering
  \setlength\tabcolsep{1.8pt}
    \begin{tabular}{|p{3.8em}|p{7em}|p{7em}|p{6em}|}
    \hline
    \multicolumn{1}{|r|}{} & \textbf{Hand Part} & \textbf{Location} & \textbf{In networks} \\
    \hline
    \textit{2-class} & Kick - K (heel),\newline Non-Kick - NK (all others) & \multicolumn{1}{l|}{None} & \textit{TablaCNN,\newline PercCNN,\newline PercVAE} \\
    \hline
    \textit{4-class} & Heel - H, Thumb - T, Fingers - F, Nails - N & None  & \textit{TablaCNN,\newline PercCNN,\newline PercVAE} \\
    \hline
    \textit{4-class + 5-class (Hierarchical)} & Heel - H, Thumb - T, Fingers - F, Nails - N & Soundhole, Upper Bout, Lower Bout, Upper Side, Lower Side & \textit{TablaCNN,\newline PercCNN} \\
    \hline
    \end{tabular}%
    \caption{Output layers mapped to guitar body percussion taxonomy.}
  \label{tab:netoutput}%
\end{table}%

\begin{figure}
  \centerline{
  \includegraphics[width=1.1\columnwidth]{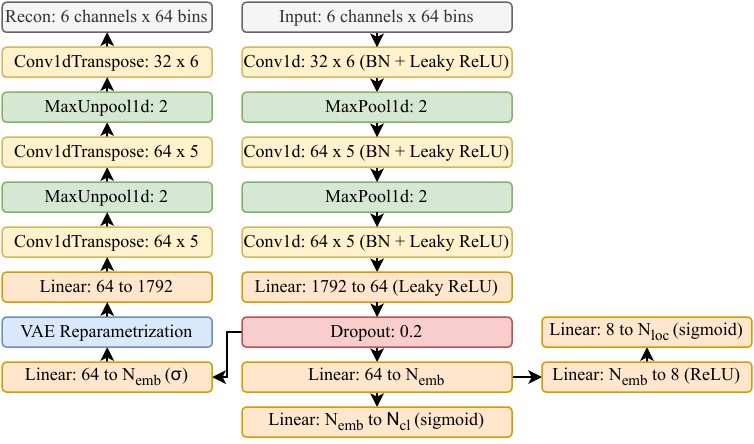}}
  \caption{Architecture of \emph{PercCNN}. The extra layers for
  location classification are on the right-hand side, the decoder
  of \emph{PercVAE} on the left. \(N_{emb}\ = 2\), \(N_{loc}\ = 5\),
  \(N_{cl}\ = 2\) or \(4\). }
  \label{fig:netarch}
 \end{figure}

\subsection{Network architectures}

\begin{table*}[ht!]
  \centering
  
  \setlength\tabcolsep{5pt}
    \begin{tabular}{|l||r|r|r|r|r||r||r|r|r|}
    \hline
    \multicolumn{1}{|r||}{} & \multicolumn{5}{c||}{\textit{GPercRep}} & \multicolumn{1}{c||}{\textit{GPercHeel}} & \multicolumn{3}{c|}{\textit{GPercPat}} \\
    \hline
    \multicolumn{1}{|r||}{} & \multicolumn{1}{l|}{\textbf{K}} & \multicolumn{1}{l|}{\textbf{NK}} & \multicolumn{1}{l|}{\textbf{F}} & \multicolumn{1}{l|}{\textbf{N}} & \multicolumn{1}{l||}{\textbf{W/Avg}} & \multicolumn{1}{l||}{\textbf{Recall}} & \multicolumn{1}{l|}{\textbf{K}} & \multicolumn{1}{l|}{\textbf{NK}} & \multicolumn{1}{l|}{\textbf{W/Avg}} \\
    \hline
    \hline
    \textbf{TablaCNN - 2-class} & 97.46 & 99.42 &       &       & 99.05 & 85.69 & 44.44 & 85.07 & 74.56 \\
    \hline
    \textbf{PercCNN - 2-class} & \textbf{98.33} & \textbf{99.61} &       &       & \textbf{99.37} & \textbf{91.68} & 0.00 & 85.14 & 63.10 \\
    \hline
    \textbf{PercVAE- 2-class} & 97.87 & 99.51 &       &       & 99.20 & 85.02 & 0.00 & 85.14 & 63.10 \\
    \hline
    \hline
    \multicolumn{1}{|r||}{} & \multicolumn{1}{l|}{\textbf{H}} & \multicolumn{1}{l|}{\textbf{T}} & \multicolumn{1}{l|}{\textbf{F}} & \multicolumn{1}{l|}{\textbf{N}} & \multicolumn{1}{l||}{\textbf{W/Avg}} &  &  &  & \\
    \hline
    \textbf{TablaCNN - 4-class} & \textbf{97.48} & \textbf{91.06} & \textbf{89.81} & \textbf{94.46} & \textbf{92.92} & 91.35 & 35.71 & 87.32 & 73.97 \\
    \hline
    \textbf{PercCNN - 4-class} & 94.61 & 78.95 & 80.86 & 93.26 & 86.92 & 69.05 & \textbf{74.29} & \textbf{93.33} & \textbf{88.40} \\
    \hline
    \textbf{PercVAE - 4-class} & 97.44 & 90.30 & 87.44 & 93.16 & 91.63 & 81.86 & 0.00 & 85.14 & 63.10 \\
    \hline
    \textbf{TablaCNN - Hierarchical} & 96.61 & 92.24 & 89.59 & 93.99 & 92.77 & 89.18 & 23.08 & 86.11 & 69.80 \\
    \hline
    \textbf{PercCNN - Hierarchical} & 95.73 & 82.05 & 87.60 & 94.18 & 90.12 & 69.55 & 0.00 & 85.14 & 63.10 \\
    \hline
    \end{tabular}%

    \caption{F-Measure (as percent) for each network on the three test datasets
    (hold-out of \emph{GPercRep}, \emph{GPercHeel} and
    \emph{GPercPat}). \textbf{H} = heel, \textbf{K} = kick,
    \textbf{T} = thumb, \textbf{NK} = non-kick, \textbf{F} = fingers,
    \textbf{N} = nails, \textbf{W/Avg} = weighted average.}
  \label{tab:classification}%
\end{table*}%

The baseline model for our experiments is an adaptation of the tabla
transcription model proposed in MA \emph{et al.}\cite{mafour}. This network processes
three stacked spectrograms with different time/frequency resolutions
on a window of 150 ms. Each frame of the
spectrogram has an 80-bin Mel representation of a window.

To adapt this network to real-time requirements we constrained
the input window to be 512 samples,
or 11.6 ms. Our adaptation (\emph{TablaCNN})
receives one window of six single Mel-frequency spectra,
one for each pickup of the prototype.
A further modification (\emph{PercCNN}) processes down-sampled FFT
features through three one-dimensional convolutional
layers
and a bottleneck layer of two dimensions before the output (Figure \ref{fig:netarch}).
To perform dimensionality reduction jointly with classification,
we implemented reparametrisation from the bottleneck layer and
a decoder mirroring the encoding CNN (\emph{PercVAE}).

\subsection {Output classes}

The labels according to the guitar body taxonomy were simplified to:
(i) a 2-class scenario with ``kicks'' (heel hits, in reference to kick drum sounds that heel
hits are supposed to imitate) and ``non-kicks''; a 4-class output
implementing all hand parts; (ii) 4-class hand part plus another 5-class
output trained on hit location on the body (hierarchical output).
The hierarchical output
was not implemented on the VAE.
This gave us a total of eight network configurations.
Tables \ref{tab:netarch} and \ref{tab:netoutput} illustrate differences between
network architectures, and the mappings between the taxonomy in Section \ref{eq:taxonomy}
and the output layers.

Loss functions used were Binary Cross-Entropy for 2-way classification,
Cross-Entropy for 4-way classification, and a sum of two equally weighted
Cross-Entropies for
hierarchical classification (hand part and location).
The VAE used the following loss function, where
\(\gamma = 0.001\) and \(\beta = 3\) after hyperparameter search,
and BCE replaced by Cross-Entropy in the four-class model: \\
\(L_{VAE} = BCE + \gamma (MSE_{Recon} + \beta KLD)\)


\subsection{Training, data augmentation, cross-validation}

All networks were trained on the \emph{GPercRep}
dataset with hold-out cross-validation:
a stratified 20\% of the
shuffled dataset was reserved for testing,
whereas the remainder of the examples was used for
training (80\%) and validation (20\%). All networks
were trained with an Adam optimiser for 100 epochs with
a batch size of 128, saving the model with the highest accuracy
on the test set. We trained with the following
data augmentation functions: high-pass at 80 Hz, high-pass at 160 Hz, \(tanh()\)
waveshaping distortion with gain of 5, phase inversion and
random changes in gain between the six channels of each
example. Those functions are meant to represent the different
input impedances and different gains of other audio preamplifiers.

Further to the \emph{GPercRep} dataset, generalisation was
checked by performing cross-dataset evaluation
\cite{inproceedings}. We recorded a snippet of real-world
guitar percussion patterns (\emph{GPercPat}); musically coherent patterns,
still with no tonal sounds, were played, rather than hits
repeated every second. Acquisition was done with
a different audio interface,
which led to a different combination of gains and frequency
responses across channels in the input audio. The dataset was annotated only with
a ``kick" and ``non-kick'' label,
leading to 85 hits in one minute of audio.

Testing on the \emph{GPercPat}
dataset highlighted a bias in our networks against
heel hits or ``kicks''. An explanation was thought to be the lack
of balance across classes in the dataset, however the issue was
not mitigated by balancing the dataset, ensuring the same number
of examples for each category. To gather further information about
this phenomenon, we created a third dataset
consisting exclusively of 601 heel hits, acquired and labelled
with the same taxonomy as \emph{GPercRep}: this will be called
\emph{GPercHeel}.


\subsection{Evaluation metrics}\label{sec:evalembeddings}

The classification performance of the networks was evaluated with Precision,
Recall and F-measure for each category,
as a 2-class or 4-class problem (see Table \ref{tab:netoutput}).

We also wanted to quantitatively
investigate the quality of the network's embeddings.
Thus, we made subsets of the data in \emph{GPercRep}
according to each label
the network was trained on (kick VS non-kick or
hand part), and for each category, we drew distributions
for each of the other parts in \emph{GPercRep}'s
taxonomy: for example, we divided non-kicks
according to their location or their dynamics.
Then we calculated the KL-divergences between the probability distributions
of each sub-category. The hypothesis behind this method
is that, if the embeddings do not carry any meaningful
information beyond the classes that the network was trained on,
the distributions will overlap and their KL-divergences
will be small and noisy. If, on the other hand, different hit
properties lead to different positions in the embeddings,
KL-divergences will be different across sub-categories and
the sub-categories will be arranged following a certain order
of similarity.

Reconstruction metrics for the VAE were not evaluated beyond their
inclusion in the loss function.
Future work could focus on the correlation between better reconstruction
and better separation of each feature.

\section{Evaluation}\label{sec:results}

\begin{figure*}
  \centerline{
    \begin{subfigure}{.33\textwidth}
      \subcaption{\emph{PercCNN}}
      \includegraphics[width=\textwidth]{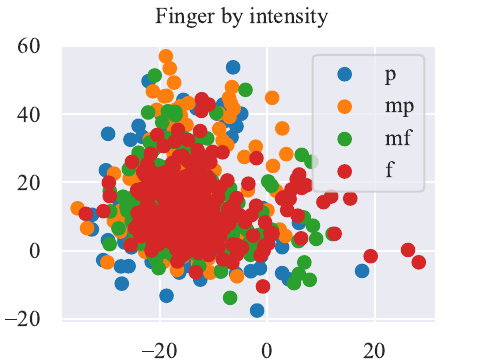} \\
      \includegraphics[width=\textwidth]{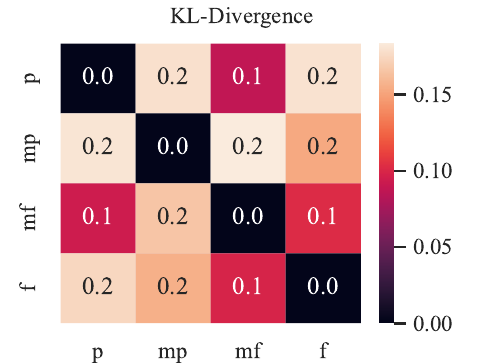}
      \label{fig:embedperc}
    \end{subfigure}
    \begin{subfigure}{.33\textwidth}
      \subcaption{\emph{TablaCNN}}
      \includegraphics[width=\textwidth]{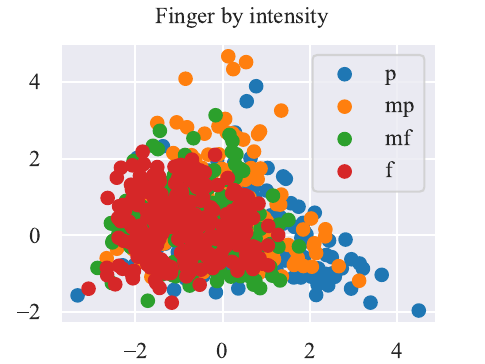} \\
      \includegraphics[width=\textwidth]{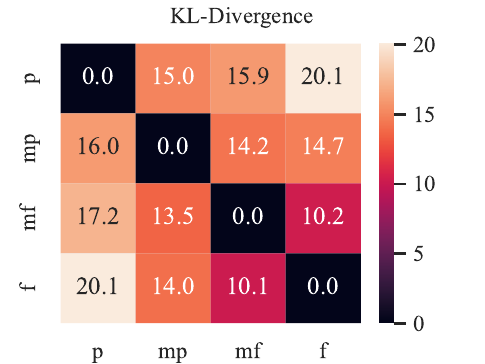}
      \label{fig:embedhier}
    \end{subfigure}
    \begin{subfigure}{.33\textwidth}
      \subcaption{\emph{PercVAE}}
      \includegraphics[width=\textwidth]{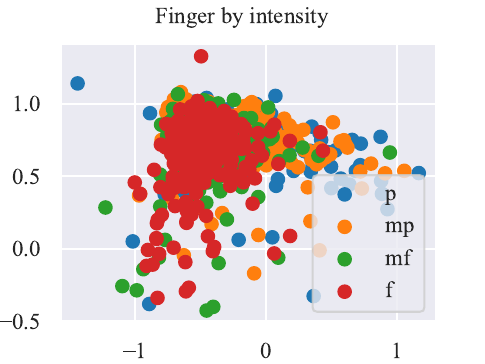} \\
      \includegraphics[width=\textwidth]{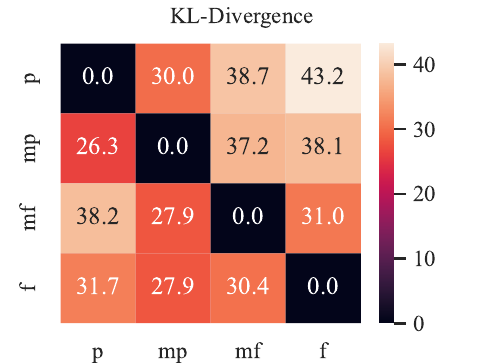}
      \label{fig:embedvae}
    \end{subfigure}
  }
  \caption{Embeddings from \emph{GPercRep}: example with
  finger hits labelled by dynamics, with matrix of KL divergence across
  the distributions of each dynamic level.}
  \label{fig:embedplots}
\end{figure*}

\subsection{Classification}

Table \ref{tab:classification} contains the F-Measure for
the predictions of each network,
with the three test datasets. In the case of
\emph{GPercHeel}, only the Recall is reported;
the Precision is always 1, as all hits are heel hits
and there cannot be false positives (non-heel hits
classified as heel hits).

\textbf{2-Class discrimination.} All networks are able to precisely
discriminate between kicks and non-kicks with an F-measure
above 99\%.
The \emph{GPercHeel} dataset shows much reduced but still
effective classification, especially with \emph{PercCNN}.
However, the test on \emph{GPercPat} exposes a generalisation problem:
despite performing data augmentation during
training, all networks show a bias toward non-kicks.
\emph{PercCNN} and \emph{PercVAE} return only non-kicks in the
dataset, despite the two classes being visually separable
when data points are extracted and plotted from their embeddings
(not pictured). This result may suggest that the
networks still overfit to the extent that they are
very sensitive to the way that the data is acquired.

\textbf{4-class discrimination.} Uniformly across the tests,
the networks yield an F-measure around 90\% for 
\emph{GPercRep}. The introduction of the classification by
location (in the two \emph{Hierarchical} networks) does not affect
the score of the hand-part classifier.
\emph{GPercHeel} yields a similar Recall
score, although higher in the case of \emph{TablaCNN}
networks. Interestingly, the weakest model in
\emph{GPercRep}, the 4-class \emph{PercCNN},
ends up being the best model in \emph{GPercPat},
although an F-measure of 74\% for kick hits may still
not be satisfactory in musical
performance. \emph{PercVAE} shows better performance
than \emph{PercCNN}, although it still fails to generalise to
\emph{GPercPat} and defaults to flagging all events as non-kicks.

The F-Measures in our results
are higher on average than the ones found in MA \textit{et al.}'s
work on tabla hit transcription \cite{mafour}.
At the same time, our
results fall below the 95\% accuracy achieved by Stefani \cite{stefaniComparisonDeepLearning2022} with an
8-class discriminator on guitar techniques, and the 97\%
by Jathal \cite{jathalRealTimeTimbreClassification2017} on the three-way discriminator for
tabletop drumming. These results, however, are not directly comparable
as the figures refer to different datasets.

\subsection{Computation times and latency}\label{sec:latencymeas}

\begin{table}[h!]
  \begin{center}
    \begin{tabular}{|l|r|r|r|r|}
      \hline
            & \multicolumn{1}{l|}{\textbf{Avg}} & \multicolumn{1}{l|}{\textbf{Std Dev}} \\
      \hline
      \textbf{PercCNN} & 0.496 & 0.332 \\
      \hline
      \textbf{TablaCNN} & 0.422 & 0.496 \\
      \hline
      \textbf{PercCNN} in Max & 12.675 & 1.132 \\
      \hline
      System (\textbf{PercCNN}) & 22.310 & 0.670 \\
      \hline
      System (no NN) & 9.922 & 0.020 \\
      \hline
      \end{tabular}%
 \end{center}
  \caption{Computation times in \(\mu\text{s}\) of both networks measured through
  TorchScript in a C++ wrapper, then end-to-end within Max and with an
  external analogue excitation.}
  \label{tab:latencymeasure}
 \end{table}

Our models all require a fixed 11.6 ms input buffer to populate the input window
after an event is detected, for example through a time-based attack detector
\cite{lucaturchetHardRealTimeOnset2018}. TorchScript was used to wrap the two-class (\emph{PercCNN} and
\emph{TablaCNN}) into a C++ test routine and a Max/MSP external
for a synthetic soak test and real-world latency tests
on a laptop with an Intel i7-8665U CPU running Windows
(Table \ref{tab:latencymeasure}).

\emph{PercCNN} and
\emph{TablaCNN} have comparable latencies in the synthetic test.
They both execute in less
than half a millisecond on average when called 10,000 times.
The real-world latency measured
manually within Max/MSP (over 30 examples) reports a value that is consistent with the 11.6 ms
window plus the synthetic timing reported above, with the attack detection
not introducing much further latency or jitter. The low-power laptop
used requires an audio buffer size of 256 samples to comfortably run \textbf{PercCNN}
in real time alongside a suitable synthesis engine:
the total system latency jumps to 22 ms when probing with a
Bela\footnote{\texttt{\url{https://bela.io}}}
board attached to the laptop's sound card (averaged over 500 examples).
Input and output buffers can be greatly reduced on ad-hoc hardware or software.


\subsection{Embeddings}


As introduced in Section \ref{sec:evalembeddings},
the distribution of subclasses of the
taxonomy within each class was explored in the
embeddings of each network. In addition to a visual
and qualitative inspection of the distribution through
scatter plots, KL-Divergence is used here as a similarity metric
to measure the distance between distributions.
In the following analysis, we will take finger hits
divided according to dynamics as an example,
but our observations are valid
for all other hand parts, and versus hit location
(e.g. heel hits divided by body location).

\textbf{Classifier embedding.} When \emph{PercCNN} is only trained as a classifier,
the four dynamic points overlap in the embeddings (Figure
\ref{fig:embedperc}). KL-Divergences range between
\(2 \cdot 10^{-5}\) and 0.2, so this range of numbers will
be used as a baseline for the interpretation of further results.
\emph{PercCNNHierarch}, trained to discriminate
according to hand part and location,
shows very precise segmentation of hit
locations but no meaningful segmentation by dynamics
(not pictured\footnote{
All pictures of embeddings available at
\url{https://github.com/iamtheband/martelloni_et_al_ismir2023}
}).

\textbf{Embedding with PCA.} \emph{TablaCNN}'s embeddings are not a
bottleneck within the network itself,
but they are calculated through PCA on the 128-dimensional dense layer.
Principal Component Analysis is shown to disentangle some
of the other features in the dataset, as the dynamics
subclasses are distributed
along a right-to-left gradient (Figure \ref{fig:embedhier}).
The KL-Divergence across those distributions
reaches a maximum of 20.1.

\textbf{Embedding/VAE latent space.} \emph{PercVAE}
shows a similar but more pronounced subdivision in the 2-dimensional
latent space. The right-to-left gradient is
visible but the KL-Divergence is much greater at a maximum of 43.2.
The KL-Divergence values steadily increase from \emph{p} to \emph{f},
more evidently than in the embeddings extracted via PCA.
This was noticeable also when
hits were segmented by location (not pictured):
for example, Lower Side
had a KL-Divergence of 30.8 versus Upper Side, 31.7 vs Lower Bout,
38.7 vs Upper Bout, and 40.7 vs Soundhole.

\section{Discussion}\label{sec:discussion}

The evaluation shows that all our models act as very accurate
2-class classifiers. Even though classification accuracy
is not as high as in other situations (with different datasets),
the simplicity of our models and the 11.6 ms input buffer
makes them faster than those systems, and well suited for
implementation on an edge device.

\textbf{Challenges.} The main issue arising from our evaluation
is the poor generalisation to our \emph{GPercPat} dataset.
Still, we have anecdotal
evidence that these networks do not behave like poor
classifiers in the real-world context of musical performance
with our augmented guitar prototype, the HITar\footnote{
Performance of the HITar at the Guthman Musical Instrument Competition 2023:
\texttt{\url{https://www.youtube.com/live/NPtHGYH0JV0?t=1150}}

HITar's Linktree: \texttt{\url{https://linktr.ee/hit4r}}
}.
The 2-class \emph{PercCNN} was coupled with a time-domain hit
detector and made to run in
real time; its continuous output probability
was mapped to a linear interpolation of parameters
on the modal synthesis engine
MetaSynth by CNRS-AMU PRISM \cite{conanNavigatingSpaceSynthesized2013};
the signal chain was connected to a different guitar (same make and model)
to the one the network was trained with; the network is able to
reliably adapt synthesis parameters even when used by players other
than the main author.
There is scope to expand the training and the evaluation by
involving more guitars, more players and different data augmentation
techniques. However, the augmented guitar that we built allows
us to pursue a further type of \emph{behavioural} evaluation with
guitar players. In particular,
musicians performing in real time
may adapt their gestures until they reliably
produce a desired set of outcomes, something not possible with
pre-recorded data. A study on the performance of guitar players
with different network configurations running on the augmented
guitar prototype will help investigate the degree to which
the musicians can adapt to the expectation of the network;
such a study would
continue our work in \cite{NIME21_32}.

\textbf{Support for rich interaction.} We observed
that \emph{PercVAE} is able to encode differences
in hit dynamics and location within the embeddings without being
trained to discriminate between them; rather than separating them
with decision boundaries like \emph{PercCNNHierarch},
each subcategory overlaps neighbouring subcategories, providing a
smooth transition that could map well to continuous quantities
such as dynamics or location on a surface.
The use of a bottleneck layer is also a more efficient solution
than PCA, as performing 
PCA would require extra matrix computation that was not captured
in the timings measured at Section \ref{sec:latencymeas}.
The parameters of a synthesis engine such as MetaSynth could be controlled not just
by the
categorical output of the discriminator, but also by the latent
representation of the VAE, either directly or through a transform.
A mapping function could be designed between the embeddings and synthesis
parameters, or the embedding vectors could be exposed directly to
synthesisers as MIDI Polyphonic Expression (MPE) \cite{romoMIDIStandardMusic2018}
controls.

\section{Conclusions}

We presented three adaptations of Automatic
Drum Transcription for guitar body percussion classification and
embedding learning, to support real-time music performance and the
augmentation of an acoustic guitar through Deep Neural Networks.
We chose and simplified a model
for ADT that was shown to be effective in the detection of
tabla strokes; a variant was also proposed which supports high-level
continuous feature representation through the use of embeddings
jointly trained as a Variational Autoencoder's latent space.
All network configurations were trained on a dataset of percussive
fingerstyle hits acquired \emph{ad hoc}, and they were tested on a hold-out
portion of that dataset plus two other datasets of similar material.
The networks performed very well on a simplified 2-class discrimination,
and comparably to the state of the art
on the full 4-class stroke classification with smaller latency.
However, they generalise poorly on a dataset that was recorded with
different computer equipment.
The embeddings were analysed both qualitatively and quantitatively through
KL-Divergence between subclasses in the taxonomy;
they show that the network encodes some information beyond the categories
with which it was trained. We argue that this information
can be used to support richness in musical interaction with
digital and augmented instruments based on DNN analysis.

\newpage

\section{Acknowledgements}

This work was supported by UK Research and Innovation's
CDT in AI \& Music [grant number EP/S022694/1]
and by PRISM Laboratory (CNRS, Aix-Marseille University).

\bibliography{bibliography}

%
%
%
%
%

\end{document}